\documentclass[aps,pra,showpacs,twocolumn,superscriptaddress]{revtex4}
\usepackage{graphicx,color}
\usepackage{amsmath,amsthm,amsfonts,amssymb,bm}
\usepackage{times}
\usepackage{epsf}
\usepackage[colorlinks={true}]{hyperref}

\usepackage[T1]{fontenc}
\usepackage[utf8]{inputenc}
\usepackage{epstopdf}

\hypersetup{citecolor={blue}, filecolor={blue}, linkcolor={blue}, urlcolor={blue}}

\newcommand{\commentold}[1]{}
\DeclareMathSymbol{:}{\mathpunct}{operators}{"3A}

\begin{document}
\title{Time-invariant Discord: High Temperature Limit and Initial Environmental Correlations}
\author{F. T. Tabesh}
\affiliation{Department of Physics, University of Kurdistan, P.O.Box 66177-15175, Sanandaj, Iran}
\affiliation{Turku Center for Quantum Physics, Department of Physics and Astronomy, University of Turku, FIN-20014 Turku, Finland}
\author{G. Karpat}
\affiliation{Faculdade de Ci\^encias, UNESP - Universidade Estadual Paulista, Bauru, SP, 17033-360, Brazil}
\author{S. Maniscalco}
\email{smanis@utu.fi}
\affiliation{Turku Center for Quantum Physics, Department of Physics and Astronomy, University of Turku, FIN-20014 Turku, Finland}
\affiliation{Center for Quantum Engineering, Department of Applied Physics, Aalto University School of Science, P.O. Box 11000, FIN-00076 Aalto, Finland}
\author{ S. Salimi}
\email{shsalimi@uok.ac.ir}
\affiliation{Department of Physics, University of Kurdistan, P.O.Box 66177-15175, Sanandaj, Iran}
\author{ A. S. Khorashad}
\affiliation{Department of Physics, University of Kurdistan, P.O.Box 66177-15175, Sanandaj, Iran}
\date{\today}

\begin{abstract}
We present a thorough investigation of the phenomena of frozen and time-invariant quantum discord for two-qubit systems independently interacting with local reservoirs. Our work takes into account several significant effects present in decoherence models, which have not been yet explored in the context of time-invariant quantum discord, but which in fact must be typically considered in almost all realistic models. Firstly, we study the combined influence of dephasing, dissipation and heating reservoirs at finite temperature. Contrarily to previous claims in the literature, we show the existence of time-invariant discord at high temperature limit in the weak coupling regime, and also examine the effect of thermal photons on the dynamical behaviour of frozen discord. Secondly, we explore the consequences of having initial correlations between the dephasing reservoirs. We demonstrate in detail how the time-invariant discord is modified depending on the relevant system parameters such as the strength of the initial amount of entanglement between the reservoirs.
\end{abstract}

\maketitle

\section{Introduction}

Quantum theory is undoubtedly a cornerstone of modern physics and one of the pillars of all natural sciences. Even though it is a century old, some of its counter intuitive features have been started to be exploited on a fundamental level, only for the last two decades, with the emergence of the quantum information science and quantum computation theory \cite{nielsenbook}. Quantum algorithms and quantum communication protocols have the potential to offer tremendous advantage over their corresponding classical counter parts. Genuine quantum correlations among the constituents of quantum systems are considered as the main resource for quantum technologies. Although the concept of quantum entanglement has been the only resource for almost all known quantum information tasks for many years \cite{entreview}, several other quantifiers of genuine quantum correlations have been introduced in recent years. Among them, quantum discord has stood out and been extensively studied in the recent literature for its role as a significant alternative resource for quantum technologies \cite{discordreview}. However, quantum systems are extremely fragile in real world conditions as they tend to rapidly lose their characteristic quantum features, such as quantum correlations, and become classical by uncontrollably interacting with their environment \cite{openbook}.

Therefore, one of the major challenges for the practical implementation of quantum technologies is the development of reliable methods to retaliate or avoid the destructive effects of this unavoidable system-environment interaction. One way of protecting the precious quantum correlations in the system is to actively modify the properties of quantum processes, through the use of various methods such as dynamical decoupling techniques, to counter the effects of decoherence \cite{dd1,dd2,dd3}. On the other hand, an alternative strategy is to initiate our system in an appropriate state depending on the properties of the system-environment interaction such that quantum correlations in the system become frozen for a certain time interval despite the detrimental effects of the environment. It has been demonstrated both theoretically \cite{frozthry1} and experimentally \cite{frozexp1,frozexp2} that, under a suitable setting, quantum correlations quantified by quantum discord might become frozen while classical correlations keep decaying until a critical time is reached. At this point, a sudden transition takes place, classical correlations freeze and quantum discord begins to diminish, giving rise to a curious phenomenon known as the sudden transition between classical and quantum decoherence.

Frozen quantum discord has been first shown to occur in a Markovian pure dephasing model, where a bipartite system interacts with two independent environments, for a family of Bell-diagonal initial states \cite{frozthry1}. Later on, it has been found out that multiple intervals of frozen discord can emerge during the dynamics under a non-Markovian random telegraph noise setting \cite{frozthry2}. Even more remarkably, it has been revealed for a dephasing model with an ohmic type spectral density that quantum discord can in fact become forever frozen throughout the time evolution of the system, that is, no sudden transition point exists and discord becomes time-invariant, remaining constant at its initial value at all times \cite{invdisc}. The existence of time-invariant discord is known to be inherently related with the emergence of non-Markovian memory effects.

Even though a lot of effort has put into the exploration of frozen and time-invariant discord phenomena, possible effects of a general non-Markovian noise setting, including the influences of dephasing, dissipation and heating channels, has not been investigated in the literature. Another unexplored problem is related to the consequences of having initial correlations between the environments for the occurrence of frozen and time-invariant discord. In other words, the main questions that we aim to answer in this work are the following: How are the properties of frozen discord affected by the presence of thermal reservoirs and what is the influence of thermal photons on this phenomenon? How is the frozen behaviour of quantum discord modified in case we initially have correlations between the environments? Specifically, here we will consider bipartite quantum systems initially prepared in Bell-diagonal states. Studying two different decoherence models, i.e., firstly a quite general combination of independent non-Markovian two-qubit dephasing, dissipation and heating channels at the high temperature limit, and secondly a two-qubit dephasing channel with initial environmental correlations, we will extensively explore the characteristics of the frozen and time-invariant discord phenomena. Our results indeed reveal the actual behaviour of time-invariant and frozen discord in decoherence models having realistic effects.

This paper is structured as follows. The concept of quantum discord is introduced in Sec. II. The time-invariant discord is investigated in the high temperature limit in Sec. III and the effects of various parameters of the environment on its existence is discussed. Sev. IV deals with the possible influences of initial environmental correlations on the behaviour of time-invariant discord. Sec. V includes the summary of our outcomes and our conclusion.

\section{Quantum Discord}

Before investigating some of the remarkable dynamical properties of quantum discord under decoherence channels, let us first briefly introduce its definition. The sum of classical and quantum correlations present in a bipartite quantum state can be measured with the help of quantum mutual information
\begin{equation}
I(\rho^{AB})=S(\rho^{A})+S(\rho^{B})-S(\rho^{AB}),
\end{equation}
where  $\rho^{k}$ $(k=A, B)$ and $\rho^{AB}$ are respectively the reduced density matrices of the subsystems and the total system, with $S(\rho)=-\textmd{Tr}(\rho \textmd{log}_{2} \rho)$ being the von-Neumann entropy. Classical correlations in a bipartite quantum system, on the other hand, can be quantified as \cite{disc1}
\begin{equation}
\emph{C}(\rho^{AB})=\sup_{\{\Pi_{k}\}}(S(\rho^{A})-S(\rho|\{\Pi_{k}\})),
\end{equation}
where the optimization is evaluated over all projective measurements $\{\Pi_{k}\}$, performed locally on the subsystem $B$. The quantum conditional entropy with respect to this measurement is then given by $S(\rho|\{\Pi_{k}\})=\sum_{k} P_{k}S(\rho_{k})$, where the conditional density operator $\rho_{k}$ associated with the measurement result $k$ can be written as
\begin{equation}
\rho_{k}=\frac{(I\otimes\Pi_{k})\rho(I\otimes\Pi_{k})}{P_{K}}
\end{equation}
with the probability $P_{k}=Tr[(I \otimes\Pi_{k})\rho(I\otimes\Pi_{k})]$. The amount of genuine quantum correlations quantified by the quantum discord can then be expressed as \cite{disc2}
\begin{equation}
D(\rho^{AB})=I(\rho^{AB})-C(\rho^{AB}).
\end{equation}

Although it can be in general a very difficult task to analytically calculate quantum discord, for the bipartite quantum states that we will consider in this work, analytical expressions are available. In particular, assuming we have a bipartite density matrix in the following X-shaped form
\begin{equation}
\rho=\begin{pmatrix}
    a & 0 & 0 & d \\
    0 & b & w & 0 \\
    0 & w & b & 0 \\
    d & 0 & 0 & d \\
\end{pmatrix},
\end{equation}
where the off-diagonal terms are real numbers, quantum discord can be evaluated analytically \cite{discformula}. It is given by
\begin{equation} \label{discord}
D(\rho)=\min\{D_{1},D_{2}\},
\end{equation}
where the minimum is simply calculated for the terms
\begin{align}
D_{1} =& S(\rho^{A})-S(\rho^{AB})-a \log_{2}(\frac{a}{a + b})-b \log_{2}(\frac{b}{a+b}) \nonumber \\
      -&b \log_{2}(\frac{b}{d+b})-d \log_{2}(\frac{d}{d+b}),     \nonumber
\end{align}
\begin{align}
D_{2}=&S(\rho^{A})-S(\rho^{AB})-\Delta_{+}log_{2}\Delta_{+}-\Delta_{-}log_{2}\Delta_{-},\nonumber
\end{align}
where $\Delta_{\pm}=\frac{1}{2}(1\pm M)$ and $M^{2}=(a-d)^{2}+4(|z|+|w|)^{2}$.

\section{Time-invariant Discord at the High Temperature Limit}
We commence this section by introducing a heuristic model where a qubit is coupled to a composite reservoir including the effects of dephasing, dissipation and heating \cite{model1,model11}. Let us consider the following time-local master equation
\begin{eqnarray} \label{master}
&&\frac{d\rho}{dt}=\frac{-i}{2}(\omega+h(t))[\sigma_{z},\rho]+\frac{\gamma_{z}(t)}{2}[\sigma_{z}\rho\sigma_{z}-\rho]\nonumber\\
&&\hspace{8mm}+\frac{\gamma_{1}(t)}{2}(\sigma_{+}\rho\sigma_{-}-\frac{1}{2}\{\sigma_{-}\sigma_{+},\rho\})\nonumber\\
&&\hspace{8mm}+\frac{\gamma_{2}(t)}{2}(\sigma_{-}\rho\sigma_{+}-\frac{1}{2}\{\sigma_{+}\sigma_{-},
\rho\}),
 \end{eqnarray}
with $\sigma_{\pm}$ being the raising and lowering operators of the qubit, $\sigma_z$ the Pauli spin operator in the z-direction, $\omega$ the transition frequency of the qubit, $h(t)$ a time-dependent frequency shift, and $ \gamma_{1,2,z}$ time-dependent decay rates. While the first term describes the Lamb shift corrections to the free Hamiltonian, the second, third and the fourth terms describe dephasing, heating, and dissipation, respectively. The state of the qubit at time $t$ can then be written as $\rho(t)=\Lambda_{\omega}(t)\rho(0)$, where $\rho(0)$ is the initial state and $\Lambda_{\omega}(t)$ is given by
\begin{equation}
 \Lambda_{\omega}(t)=\begin{pmatrix}
    1 & 0 & 0 & 0 \\
    0 & \eta_{\perp}(t)\cos(\phi(t)) &  -\eta_{\perp}(t)\sin (\phi(t))& 0 \\
    0 &  \eta_{\perp}(t)\sin (\phi(t)) &  \eta_{\perp}(t)\cos(\phi(t))&0 \\
   \kappa(t) & 0 & 0 &  \eta_{\parallel}(t) \\
\end{pmatrix},
\end{equation}
where the elements of the above matrix $\Lambda_{\omega}(t)$ read
\begin{eqnarray} \label{ele}
&&\phi(t)=\omega t+\theta(t),\nonumber\\
&&\eta_{\parallel}=e^{-\Gamma(t)},\nonumber\\
&&\eta_{\perp}=e^{-\Gamma(t)/2-\Gamma_z(t)},\nonumber\\
&&\kappa(t)=-e^{-\Gamma(t)}(1+2G(t))+1,
\end{eqnarray}
and the four terms appearing in Eq. (\ref{ele}) are expressed as
\begin{eqnarray} \label{ele2}
&&\theta(t)=\int^{t}_{0} d t^{\prime}  h(t^{\prime}),\nonumber\\
&&\Gamma(t)=\int^{t}_{0}dt^{\prime}\hspace{1mm}(\gamma_{1}(t^{\prime})+\gamma_{2}(t^{\prime}))/2,\nonumber\\
&&\Gamma_{z}(t)=  \int^{t}_{0} dt^{\prime}\hspace{1mm} \gamma_{z}(t^{\prime}),\nonumber\\
&&G(t)=\int^{t}_{0} d t^{\prime}\hspace{1mm} e^{\Gamma(t^{\prime})}\gamma_{2}(t^{\prime})/2.
\end{eqnarray}

In this work, we ignore the effect of the first term in Eq. (\ref{master}) and focus on a thermal reservoir at temperature $T$. We assume that the heating and the dissipation decay rates are respectively given by $\gamma_{1}(t)/2=(N)f(t)$ and $\gamma_{2}(t)/2=(N+1)f(t)$, where $N$ represents the mean number of photons in the modes of the thermal reservoir at temperature $T$. We note that, in case of a zero temperature reservoir, the heating rate vanishes, that is, $\gamma_{1}(t)=0$, and the dissipation rate is simply given by $\gamma_{2}(t)/2=f(t)$. For the model considered in our study, the spectral density is taken as $J(\omega)= \gamma_0 \lambda^2 / 2\pi[(\omega_0 - \Delta - \omega)^2 + \lambda^2]$, where $\gamma_{0}$ is an effective coupling constant which is related to the relaxation time of the system $\tau_{R}\approx1/\gamma_{0}$ and $\lambda$ is the width of the Lorentzian spectrum that is connected to the reservoir correlation time $ \tau_{B}\approx1/\lambda$. Additionally, $\Delta =\omega-\nu_{c}$ is the detuning of $\omega$ and $\nu_{c}$ is the centeral frequency of the thermal reservoir. It is worth noting that the effective coupling between the qubit and its environment decreases when the value of the detuning $\Delta$ increases \cite{coupdetun}. Taking into account these considerations, the function $f(t)$ can be written in the following form \cite{model11}
\begin{eqnarray}
&&f(t)=-2 \Re\{\frac{\dot{C}(t)}{C(t)}\},\nonumber\\
&&C(t)= e^{-(\lambda-i\Delta)t/2}(\cosh(\frac{d t}{2})+\frac{\lambda-i\Delta}{d}\sinh(\frac{d t}{2}))C(0),\nonumber\\
\end{eqnarray}
with $d=\sqrt{(\lambda-i\Delta)^{2}-2\gamma_{0}\lambda}$. We can also define $R=\gamma_{0}/\lambda$ in order to distinguish the strong coupling regime from the weak coupling regime. It has been demonstrated that in the weak coupling regime, $R\ll 1$, for sufficiently large detunings, the function $f(t)$ might take on negative values within certain time intervals, hence the dynamics of the qubit becomes nondivisible and non-Markovian \cite{nonmarkov}.

Supposing that the dephasing reservoir is at temperature $T$, then the time-dependent dephasing rate takes the form
\begin{equation} \label{gamz}
\gamma_{z}(t)=\int d\omega J(\omega)\coth(\hbar \omega/2 k_{B}T)\frac{\sin(\omega t)}{\omega}.
\end{equation}
In the high temperature limit, the above equation simply reads
\begin{equation}\label{gamz2}
\gamma_{z}(t)=\frac{2 k_{B}T}{\hbar}\int d\omega J(\omega)\frac{\sin(\omega t)}{\omega^{2}},
\end{equation}
where $\omega_{T}=k_{B}T/\hbar$ is the thermal frequency and we assume that the spectral density is of the ohmic type, i.e., $J(\omega)=\alpha(\omega^{s}/\omega^{s-1}_{c})e^{-\omega/\omega_{c}}$, with $\omega_{c}$ being the cutoff frequency, $s$ the Ohmicity parameter, and $\alpha$ the coupling constant. In addition, $\gamma_{z}(t)$ takes temporarily negative values provided $s>s_{crit}= 3$. Therefore, if we have a super-Ohmic spectral density with $s>3$, information can flow back form the environment to the system giving rise to memory effects \cite{invdisc}.

Consequently, with the help of Eqs. (\ref{ele2}-\ref{gamz2}), one can obtain the following expressions to fully describe the dynamics of a qubit coupled to the considered reservoirs,
\begin{eqnarray} \label{factors}
&&\Gamma(t)= -\Re[\ln (x(t)^{2N+1})],\nonumber\\
&&\Gamma_{z}(t)= \alpha \hspace{1mm}\frac{2k_{B}T}{\hbar \hspace{1mm}\omega_{c}} \tilde{\Gamma}(-2+s)(1-(1+\omega_{c}^{2}\hspace{1mm}t^{2})^{(-s+2)/2}\nonumber\\
&&\hspace{10mm} \times\cos((-2+s) \arctan(\omega_{c}\hspace{1mm} t))),\nonumber\\
&&\kappa(t)=\frac{-1}{2N+1}(1- \exp[-\Re[\ln (x(t)^{2N+1})]]),
\end{eqnarray}
where $x(t)=\{C(t)/C(0)\}^{2}$, $\tilde{\Gamma}(s)$ is the Euler gamma function. The memory time of the dephasing environment can be defined by $\omega^{-1}_{c}$. As a consequence, we can define $\beta=\omega_{c}/\lambda$ to characterize the relation between the cut-off frequency of the dephasing environment and the width of the Lorentzian spectrum of the thermal reservoir.

We have dealt with the description of the dynamics of a single qubit up to this point in our paper. Let us now suppose that we have a bipartite quantum system composed of two identical qubits, labelled as $A$ and $B$, that are locally coupled to their own environments. We also assume that the individual environments are identical and not correlated with each other. Hence, it is possible to obtain the time evolution of the two-qubit system from the single qubit dynamics in a straightforward fashion as $\rho^{AB}(t)= (\Lambda^{A}_{\omega}(t)\otimes\Lambda^{B}_{\omega}(t))\rho^{AB}(0)$. In the course of our work, we choose the initial state of the two-qubit open system in the form of Bell-diagonal states
\begin{equation}\label{bell}
\rho_{S}(0)=\frac{1}{4}(I\otimes I+ \sum^{3}_{i=0}m_{i}\sigma_{i}\otimes\sigma_{i}),
\end{equation}
where $m_{i}$ are three real number such that $ -1 \leq m_{i}\leq 1$, and $\sigma_i$ are the Pauli spin operators in the x,y and z directions. Therefore, the time-evolution of our system is given by
\begin{equation}
\rho_{S}(t)=\begin{pmatrix}
    \rho_{11} & 0 & 0 &  \rho_{14} \\
    0 &  \rho_{22} &  \rho_{23}& 0 \\
    0 &   \rho_{23} &   \rho_{33}&0 \\
   \rho_{14} & 0 & 0 &   \rho_{44} \\
\end{pmatrix},
\end{equation}
where the density matrix elements can be evaluated as
\begin{eqnarray}
 &&\rho_{11}=\frac{1}{4}((1+\kappa(t))^{2}+\eta^{2}_{\parallel}(t)m_{3}),\nonumber\\
 &&\rho_{22}=\rho_{33}=\frac{1}{4}(1-\kappa(t)^{2}-\eta^{2}_{\parallel}(t)m_{3}),\nonumber\\
 &&\rho_{44}=\frac{1}{4}((1-\kappa(t))^{2}+\eta^{2}_{\parallel}(t)m_{3}),\nonumber\\
 &&\rho_{23}=\frac{m_{1}+m_{2}}{4}\eta^{2}_{\perp}(t),\nonumber\\
 &&\rho_{14}=\frac{m_{1}-m_{2}}{4}\eta^{2}_{\perp}(t).
\end{eqnarray}

\subsection{Pure Dephasing}
In this subsection, we will only consider the pure dephasing case at the high temperature limit in the weak coupling regime, when the two qubits interact with independent reservoirs. We choose the initial state of our two-qubit system from a family of Bell-diagonal states, namely, from the states given by Eq. (\ref{bell}) having the three real parameters $m_{1}=1$ and $m_{2}= - m_{3}= m$, with $|m|< 1$. For such initial states and in the presence of pure dephasing dynamics, classical correlations and mutual information can be written in a compact form in the following way
\begin{align}
C(\rho^{AB})&=\sum_{j=1}^2\frac{1+(-1)^j\chi(t)}{2}\text{log}_2[1+(-1)^j\chi(t)], \\
I(\rho^{AB}(t))&=\sum_{j=1}^2\frac{1+(-1)^j m}{2}\text{log}_2[1+(-1)^j m]\nonumber\\ &+\sum_{j=1}^2\frac{1+(-1)^je^{-2\Gamma_z(t)}}{2}\text{log}_2[1+(-1)^je^{-2\Gamma_z(t)}],
\end{align}
with $\chi(t)=\max\{e^{-2\Gamma_{z}(t)}, m\}$, and thus quantum discord $D(\rho^{AB})$ can be simply evaluated from their difference. Using these equations, one can define a transition time $\tilde{t}$ as
\begin{equation} \label{cond1}
e^{-2\Gamma_{z}(\tilde{t})}= m,
\end{equation}
below which ($t <\tilde{t}$) quantum discord remains completely unaffected by the noise and classical correlations decay. On the other hand, after this critical time point ($t >\tilde{t}$), classical correlations freeze and discord starts to decrease.
\begin{figure}[t]
\includegraphics[scale=0.6]{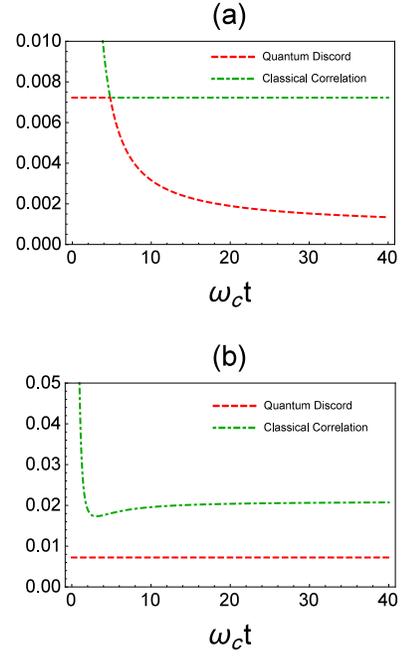}
\caption{(Color online) Dynamics of classical and quantum correlations for independent pure dephasing environments as a function of the scaled time $\omega_{c} t$ for the initial states having $m=0.1$. (a) Markovian ($s=2.5$) and (b) Non-Markovian ($s=3.5$) dynamics.}
\label{fig1}
\end{figure}
\begin{figure}[b]
\includegraphics[scale=0.5]{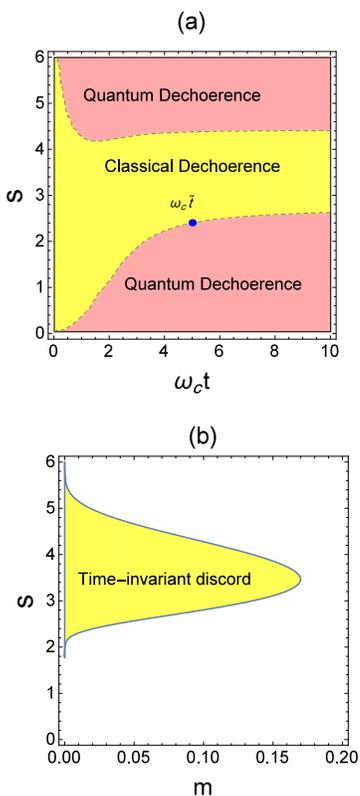}
\caption{(Color online) (a) Landscape of correlation dynamics in the $(s-\omega_{c}t)$ plane, for the initial state with $m = 0.1$ at the high temperature limit. As the pink (dark gray)area denotes the quantum decoherence regime, the yellow (gray) area shows the classical decoherence regime. (b) The yellow shaded (gray) region indicates the range of ohmicity and initial state parameters, $s$ and $m$, for which time-invariant discord exists.}
\label{fig2}
\end{figure}
Let us now assume that $\alpha=\hbar \omega_{c}/ 2 k_{B}T$ in the second line of Eq. (\ref{factors}) and hence $\Gamma_{z}(t)$ is independent of the temperature. With this consideration in mind, as we are working at the high temperature limit, i.e., $2k_{B}T\gg \hbar\omega_{c}$, the coupling between the open system and its environment will be weak.

Dynamical behaviours of the quantum discord (dashed red line) and the classical correlations (dotted-dashed green line) are plotted in Figs. \ref{fig1}(a) and \ref{fig1}(b) as a function of $\omega_{c}t$ for the ohmicity parameters $s=2.5$ and $s=3.5$, respectively, where the initial state is chosen as $m=0.1$. Since we work at the high temperature limit in the weak coupling regime, we consider that $\alpha=0.01$ and $2 k_{B}T/\hbar \omega_{c}=100$. Note that for the non-Markovian memory effects to emerge at the high temperature limit, the ohmicity parameter should satisfy the condition $s\geq3$. Looking at Fig. \ref{fig1}, we clearly see that whereas we have frozen discord for a finite time interval in case of Markovian dynamics, time-invariant discord can be observed for non-Markovian evolution. In other words, while Eq. (\ref{cond1}) has a solution for $s=2.5$, there exists no solution for it when $s=3.5$, giving rise to time-invariant discord. Therefore, we find out that the inherent connection between the non-Markovianity and the occurrence of time-invariant discord, as first explained in Ref. \cite{invdisc}, still holds at the high temperature limit. However, contrarily to what has been claimed in Ref. \cite{invdisc}, there indeed exists $s$ and $m$ even at the high temperature limit, such that the condition given in Eq. (\ref{cond1}) is never satisfied and the phenomenon of time-invariant discord can still be observed. We emphasize that the key point here leading the emergence of time-invariant discord at the high temperature limit is the fact that we are working in the weak coupling regime as no such phenomenon would be present if the coupling was strong.

In Fig. \ref{fig2}(a), we display the outlook of correlation dynamics in the presence of independent pure dephasing reservoirs in the $s-\omega_{c}t$ plane. From this plot, one can see the range of values for $s$ and $\omega_{c}t$ for which Eq. (\ref{cond1}) has a solution (frozen discord and sudden transition), and for which it does not have a solution (time-invariant discord). The intersection between the quantum and classical decoherence regions, as respectively shown by pink (dark gray) and yellow (gray) areas in the figure, correspond to the sudden transition point, after which point quantum discord begins to decay. Fig. \ref{fig2}(b) displays the asymptotic long-time limit for the transition condition given in Eq. (\ref{cond1}). Yellow shaded \textbf{(gray)} region in the figure demonstrates the area defined by the values of the ohmicity parameter $s$ and the initial state parameter $m$ for which the phenomenon of time-invariant discord exists at the high temperature limit. Outside the yellow shaded region one will always see a sudden transition from classical to quantum decoherence and thus quantum discord can be frozen only for a finite time interval.
\begin{figure}[t]
\includegraphics[scale=0.6]{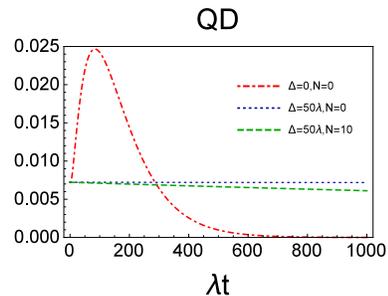}
\caption{(Color online) Dynamics of quantum discord in the presence of independent dissipation and heating reservoirs for the initial Bell-diagonal state $m=0.1$ as a function of scaled time $\lambda t$ for $R=0.01$}
\label{fig3}
\end{figure}

\subsection{Dissipation and Heating}
This section deals with the dynamics of quantum discord for two qubits independently interacting with dissipation/heating reservoirs. Using the analytical expression given in Eq. (\ref{discord}), we present a plot of the time-evolution of quantum discord as a function of $\lambda t$ in Fig. \ref{fig3}. We consider the weak coupling regime with the parameter $R=0.01$ and the initial Bell-diagonal state with $m=0.1$. We also recall that in the weak coupling regime, Markovian dynamics emerge in case of sufficiently small detuning parameter, e.g., $\Delta=0$. In the figure, the dotted-dashed red line displays the case where the dynamics is Markovian and there are no thermal photons. As can be clearly seen, although discord is initially amplified, it tends to monotonically decay and vanish in the long time limit. In fact, this behaviour can be observed for the class of initial states such that $|m|<0.2$. Moving to the non-Markovian regime with $\Delta=50 \lambda$, still in the absence of thermal photons, dotted blue line shows that, even though discord is not actually time-invariant, it decays very slowly due to energy exchange between the system and the environment, and for all practical purposes, it can be considered almost invariant. Lastly, the effect of the thermal photons is demonstrated by the dashed green line. We observe that in the short time limit quantum discord is still almost time-invariant, despite the fact that thermal photons hasten its decay in the long time limit. It is important to notice that the sudden transition does not exists under this type of noise.
\begin{figure}[b]
\includegraphics[scale=0.4]{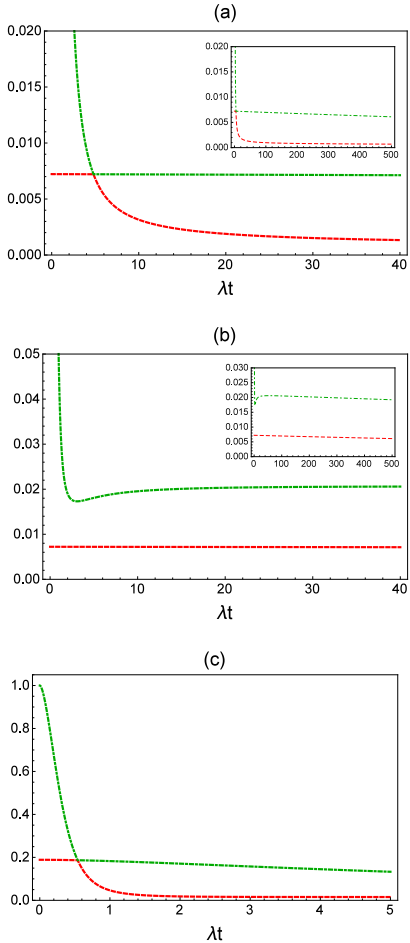}
\caption{(Color online) Dynamics of classical (dotted-dashed green line) and quantum correlations (dashed red line) for dephasing and dissipation/heating reservoirs as a function of scaled time  $\lambda t$ for $R=0.01$, $\Delta=50 \lambda$, $m=0.1$ $N=10$. (a) for $s=2.5$ and for (b) $s=3.5$. Insets are plotted for long time limits. In (c), we have $R=0.001$, $\Delta=0$, $m=0.5$, $N=10$, and $s=2.5$.}
\label{fig4}
\end{figure}

\subsection{Dephasing, Dissipation and Heating}

In this section, we take into account the combined effect of dephasing, dissipation/heating reservoirs which are once again interacting with individual qubits independently. We assume that the memory time of the dephasing and dissipation/heating environments is the same, i.e., we choose the parameter $\beta=\omega_{c}/\lambda$ to be unity. Let us also work in the weak coupling limit of both models and therefore take $\alpha=0.01$ and $R=0.01$. At the high temperature limit of the dephasing model with these conditions, and supposing detuning parameter is taken as $\Delta=50 \lambda$, the number of the thermal photons in dissipation/heating reservoir become approximately $N\approx10$.

Time evolution of the classical and quantum correlations are shown in Fig. \ref{fig4}(a) and \ref{fig4}(b) when $s=2.5$ (Markovian) and $s=3.5$ (non-Markovian) for the dephasing reservoir, respectively. We note that the effects of the dissipation/heating reservoir is also present in these plots and the dissipation/heating dynamics is non-Markovian due to the detuning parameter acquiring the value $\Delta=50 \lambda$. Thus, in Fig. \ref{fig4}(a), while the dephasing dynamics is Markovian, the dissipation/heating dynamics exhibit non-Markovian behaviour. We can observe that the sudden transition still exists even in the presence of non-Markovian dissipative environments and consequently almost time-invariant discord cannot be present. To put it differently, comparing Fig. \ref{fig3} to \ref{fig4}(a), one can see that the presence of even Markovian dephasing can dominate the dynamics over non-Markovian dissipative reservoirs in terms of time-invariant discord, causing a sudden transition and subsequent rapid decay of quantum correlations. We should stress that the quantum (classical) correlations before (after) the sudden transition point here is not actually constant but rather decay very slowly with time, which is due to the effect of dissipative reservoirs. In Fig. \ref{fig4}(b), both the dephasing and dissipation/heating dynamics are non-Markovian. This implies that the occurrence of almost time-invariant discord is exclusively related to the non-Markovian memory effects in  the purely dephasing dynamics. As can be seen in inset, it is expected that quantum discord once again degrades very slowly for long times and is not actually time-invariant. Finally, in Fig. \ref{fig4}(c), we take $R=0.001$, $\Delta=0$, $s=2.5$, $m=0.5$, and $N=10$. In other words, here both the dephasing and dissipation/heating dynamics are Markovian and there exists no memory effects. We see that the sudden transition takes place as a result of the Markovian dephasing dynamics.

\section{Time-invariant Discord in the Presence of Correlated Reservoirs}

In this section, we turn our attention to a different decoherence model that takes into account the  correlations between the environments. In particular, we consider an open system $S$ that contains two qubits interacting with a composite environment $E$, which itself is composed of two subsystems. We suppose that each qubit interacts locally with one of the environments, and $S$ and $E$ are initially not correlated, but the two environments are initially in a correlated composite state. The total Hamiltonian is given as \cite{model2}
\begin{align}
H=&H_{0}+H_{int}(t), \hspace{8mm}   H_{0}=\sum^{2}_{i=1}(H^{i}_{S}+H^{i}_{E}), \nonumber \\
H^{i}_{S}&=\epsilon_{i}\sigma^{i}_{z}, \hspace{15mm}  H^{i}_{E}=\sum_{k}\omega^{i}_{k}b^{i\dag}_{k}b^{i}_{k},
\end{align}
with $b^{i\dag}_{k}(b^{i}_{k})$ being the creation (annihilation) operator of the $k$th mode of environment $i = 1,2$, and $\sigma^{i}_{z}$ the Pauli matrix and $\epsilon_{i}$ the energy gap of the $i$th qubit. The interaction Hamiltonian is given by $H_{int}(t)=\sum_{i}H^{i}_{int}(t)$, in which local interactions are specified by
\begin{eqnarray}
&&H^{i}_{int}(t)=\chi_{i}(t)\sum_{k}\sigma^{i}_{z}\otimes(g^{i}_{k}b^{i\dag}_{k}+g^{i\ast}_{k}b^{i}_{k}),
\end{eqnarray}
where $g^{i}_{k}$ is the coupling constant between qubit $i$ and $k$th mode of its environment, $ g^{i}_{k}\in\Re$ for $i = 1,2$ and all $k$. The step function $\chi_{i}(t)$ is also given by
\begin{eqnarray}
\chi_{i}(t) =
\begin{cases}
1, & t\in[t^{s}_{i},t^{f}_{i}]\\
0, &  \textrm{otherwise}
\end{cases}
\end{eqnarray}
for some $t^{f}_{i}>t^{s}_{i}>0$. Here, the step function controls whether the local interaction of the qubit $i$ at the time $t^{s}_{i}$ and $t^{f}_{i}$ is switched on or off, respectively. Note that the duration of the local interactions can be adjusted individually for both qubits so that it is possible to study both simultaneous and consecutive interactions. Here, we will assume that $t^{s}_{1}\leq t^{s}_{2}$. Writing the interaction Hamiltonian in the interaction picture
\begin{eqnarray}
H^{I}_{int}=\sum_{j,k}\chi_{j}(t)\sigma^{j}_{z}\otimes(g^{j}_{k} e^{i \omega^{j}_{k}(t)}b^{j\dag}_{k}+g^{j\ast}_{k}e^{-i \omega^{j}_{k}(t)}b^{j}_{k}),
\end{eqnarray}
with the initially factorized state
\begin{eqnarray}
\rho(0)=\rho_{S}(0)\otimes\rho_{E}(0),
\end{eqnarray}
the time evolution of the initial state of the qubits at time $t$ in the Schr\"{o}dinger picture can be obtained as
\begin{eqnarray}
\rho_{S}(t)= Tr_{E}[U(t)\rho_{S}(0)\otimes\rho_{E}(0) U^{\dag}(t)],
\end{eqnarray}
where the propagator is given by $U(t)= e^{-iH_{0}t}U_{I}(t)$.

The initial environmental state is assumed to be a tensor product of identical two-mode Gaussian states $\eta^{k}_{12}$ of the $k$th mode of environment 1 and 2, i.e., $|\eta_{12}\rangle=\bigotimes_{k}|\eta^{k}_{12}\rangle$, hence, $\rho_{E}(0)=|\eta_{12}\rangle\langle\eta_{12}|$.
Let us suppose all two-mode Gaussian states have identical covariance matrix in standard form as
 \begin{equation}
 \textbf{S}=\begin{pmatrix}
    a & 0 & c_{+}& 0 \\
    0 &a & 0 &c_{-} \\
    c_{+} &0 & b&0 \\
   0 & c_{-} & 0 &  b \\
\end{pmatrix},
\end{equation}
and the elements of the covariance matrix are given as
\begin{eqnarray}
&&a=\frac{1}{2}\cosh(2r)+N_{1}\cosh^{2}(r)+N_{2}\sinh^{2}(r),\nonumber\\
&&b=\frac{1}{2}\cosh(2r)+N_{2}\cosh^{2}(r)+N_{1}\sinh^{2}(r),\nonumber\\
&&c_{-}=\frac{-1}{2}(1+N_{1}+N_{2})\sinh(2r)=-c_{+},\nonumber
\end{eqnarray}
with $r$ being the squeezing parameter and $N_{1,2}$ the mean occupation number in environment 1 and 2, respectively. We will here consider an ohmic type spectral density $J(\omega)=\alpha(\omega^{s}/\omega^{s-1}_{c})e^{-\omega/\omega_{c}}$ with equal cutoff frequencies $\omega_{c}$ but different coupling constants $\alpha_{j}$ for the two environments.

Let us once again assume that the initial state of the two-qubit open system can initially be described with a special family of Bell-diagonal states given by Eq. (\ref{bell}) with the three real parameters $m_{1}=1$ and $m_{2}= - m_{3}= c$, with $|c|< 1$. This is basically the same family of initial two-qubit states used for our investigation in the previous section. After some rather straightforward algebra, one can obtain the time evolution of the reduced state of the two qubits as
\begin{equation}
\rho_{S}(t)= \begin{pmatrix}
         \frac{1+c}{4}&0&0&\frac{1+c}{4}\kappa_{12}(t) \\
          0 &\frac{1-c}{4}&\frac{1-c}{4}\Lambda_{12}(t)&0 \\
          0 &\frac{1-c}{4}\Lambda^{\ast}_{12}(t)&\frac{1-c}{4}&0 \\
        \frac{1+c}{4}\kappa^{\ast}_{12}(t)&0&0&\frac{1+c}{4} \\
\end{pmatrix},
\end{equation}
where the coherence functions are given as
\begin{align}
\kappa_{12}(t) &= \kappa_{1}(t)\kappa_{2}(t)f_{1}(t)f_{2}(t)f_{3}(t)f_{4}(t),\nonumber \\
\Lambda_{12}(t)&= \kappa_{1}(t)\kappa^{\ast}_{2}(t)/(f_{1}(t)f_{2}(t)f_{3}(t)f_{4}(t)),\nonumber
\end{align}
with
\begin{align}
\kappa_{1}(t)=& e^{-2 i \epsilon_{1}t}e^{-4a \alpha_{1} \Gamma(-1+s)[2-(1-i\omega_{c}t_{1})^{1-s}-(1+i\omega_{c}t_{1})^{1-s}]},\nonumber \\
\kappa_{2}(t)=& e^{-2 i \epsilon_{2}t} e^{-4b \alpha_{2} \Gamma(-1+s)[2-(1-i\omega_{c}t_{2})^{1-s}-(1+i\omega_{c}t_{2})^{1-s}]},\nonumber \\
f_{1}(t)=&\exp\big[A(1+\omega^{2}_{c}(t_{1}+t_{2}+t^{s}_{2})^{2})^{-s/2}\times\nonumber\\
&\big(\cos[s\times\arctan(\omega_{c}(t_{1}+t_{2}+t^{s}_{2})]+\omega_{c}(t_{1}+t_{2}+t^{s}_{2})\nonumber\\
&\sin[s \times \arctan(\omega_{c}(t_{1}+t_{2}+t^{s}_{2}))]\big)\big],\nonumber \\
f_{2}(t)=&\exp\big[-A(1+\omega^{2}_{c}(t_{1}+t^{s}_{2})^{2})^{-s/2}\nonumber\\
&\times\big(\cos[s\times \arctan(\omega_{c}(t_{1}+t^{s}_{2})]+\omega_{c}(t_{1}+t^{s}_{2})\nonumber\\
&\sin[s\times \arctan(\omega_{c}(t_{1}+t^{s}_{2}))]\big)\big],\nonumber \\
f_{3}(t)=&\exp\big[-A(1+\omega^{2}_{c}(t_{2}+t^{s}_{2})^{2})^{-s/2}\nonumber\\
&\times\big(\cos[s\times \arctan(\omega_{c}(t_{2}+t^{s}_{2})]+\omega_{c}(t_{2}+t^{s}_{2})\nonumber\\
&\sin[s\times \arctan(\omega_{c}(t_{2}+t^{s}_{2}))]\big)\big],\nonumber \\
f_{4}(t)=&\exp\big[A(1+(\omega_{c}t^{s}_{2})^{2})^{-s/2}\times\nonumber\\
&\big(\cos[s \hspace{1mm}\arctan(\omega_{c}t^{s}_{2})]+\omega_{c}t^{s}_{2}\sin[s\times \arctan(\omega_{c}t^{s}_{2})]\big)\big],\nonumber
\end{align}
where $A=- 8 c_{-}\Gamma(-1+s) \sqrt{\alpha_{1}\alpha_{2}}$, the local interaction times are given by $t_{j}(t)=\int^{t}_{0}dt^{\prime}\chi_{j}(t^{\prime})$ , and $\Gamma(s)$ is the Euler gamma function. We set $t^{s}_{1}=0$ for simplicity and $t^{s}_{2}$ as the time when the interaction of the second qubit with its environment is turned on. It is important to emphasize that the free evolution of the system is taken into account in this model.

We now focus on a two-mode squeezed vacuum state that is characterized by $N1 = N2 = 0$ and defines a symmetric two-mode Gaussian state satisfying $a = b$. A squeezed vacuum state has following representation in the Fock basis \cite{fock}
\begin{align}
|\psi_{u}\rangle= \sqrt{1-u^{2}}\sum^{\infty}_{n=0}u^{n}|n\rangle\otimes|n\rangle
\end{align}
where $u = tanh(r)$ and $|n\rangle$ represents the $n$th Fock state. Notice that a squeezed vacuum state is entangled if and only if $r\neq 0$. Moreover, all two-mode Gaussian states which cannot be factorized have shown to contain non-zero genuine quantum correlations as quantified by quantum discord \cite{nonzdisc1,nonzdisc2}.

Here, in order to obtain the quantum discord for density matrix in Eq. (28) we define the projectors $\Pi_{k}=|k\rangle\langle k|$ (k = 1,2) by the orthogonal states
\begin{eqnarray}
&&|1\rangle=\cos(\theta)|\uparrow\rangle+e^{i\varphi}\sin (\theta)|\downarrow\rangle,\nonumber\\
&&|2\rangle=\sin(\theta)|\uparrow\rangle-e^{i\varphi}\cos (\theta)|\downarrow\rangle.\nonumber
\end{eqnarray}

It is not difficult to find out that the classical correlations in this case do not explicitly depend on $\varphi$ and they are maximized for $\theta = n \pi/4$ with $n \in Z$. Thus, one can obtain the analytic expression for classical correlations as
\begin{align}
C(\rho_{S}(t))= \sum^{2}_{i=1}\frac{1+(-1)^{i}\chi(t)}{2}\log_{2}(1+(-1)^{i}\chi(t))
\end{align}
where $\chi(t)=\max\{|c|, 1/2|(\kappa_{12}(t)+\Lambda_{12}(t))+c(\kappa_{12}(t)-\Lambda_{12}(t))|\}$ and we have taken c positive for the sake of simplicity. The quantum mutual information is also given by
\begin{align}
&I(\rho_{S}(t))=\sum^{2}_{i=1}\frac{1+(-1)^{i}c}{2}\log_{2}(1+(-1)^{i}c)\nonumber\\
&+\frac{1+c}{4}\sum^{2}_{i=1}(1+(-1)^{i}|\kappa_{12}(t)|)\log_{2}(1+(-1)^{i}|\kappa_{12}(t)|)\nonumber\\
&+\frac{1-c}{4}\sum^{2}_{i=1}(1+(-1)^{i}|\Lambda_{12}(t)|)\log_{2}(1+(-1)^{i}|\Lambda_{12}(t)|).
\end{align}
Then, quantum discord can be simply obtained by subtracting the classical part of correlations $C(\rho_{S}(t))$ from the total amount of correlations $I(\rho_{S}(t))$ quantified by the quantum mutual information. Based on the above equations, one can define a sudden transition time $\tilde{t}$ as
\begin{equation}\label{cond2}
1/2|(\kappa_{12}(\tilde{t})+\Lambda_{12}(\tilde{t}))+c(\kappa_{12}(\tilde{t})-\Lambda_{12}(\tilde{t}))|= c,
\end{equation}
below which ($t <\tilde{t}$) quantum discord is robust against the dephasing noise and classical correlations tend to decay. On the other hand, classical correlations remain constant for $t >\tilde{t}$, and quantum discord starts to diminish. This is naturally nothing but the same phenomenon discussed in the previous section. Depending on whether there exists a solution for Eq. (\ref{cond2}) or not, one respectively observes frozen discord accompanied by a sudden transition or time-invariant discord.

In the following, we will discuss the behaviour of the phenomena of frozen and time-invariant discord in the presence of initial environmental correlations between the two dephasing reservoirs having ohmic-type spectral densities. We will restrict our attention to the local interactions that are subsequently turned on and last for equally long times. Let us suppose that the  two qubits have the same energy gap $\epsilon_{1}=\epsilon_{2}=\epsilon=10^{-8}\omega_{c}$, and the coupling constants are taken identical, that is, $\alpha_{1}=\alpha_{2}=\alpha$. We recall that in Ref. \cite{invdisc}, in the absence of correlations between the environments, it has been demonstrated that if two qubits locally interact with pure dephasing reservoirs, frozen discord and time-invariant discord might be observed for the ohmicity parameters $s=1$ (Markovian dynamics) and $s=2.5$ (non-Markovian dynamics), respectively.
\begin{figure}[t]
\includegraphics[scale=0.6]{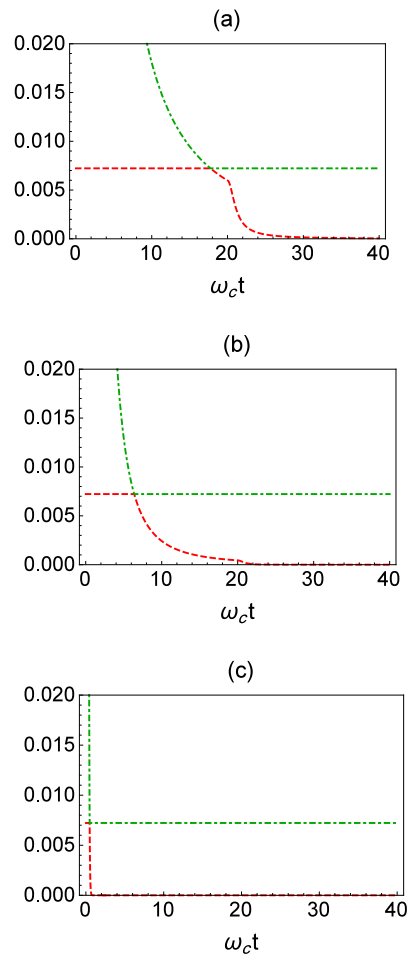}
\caption{(Color online) Dynamics of classical and quantum correlations as a function of scaled time $\omega_{c}t$. The dotted-dashed green and dashed red lines represent the classical correlations and quantum discord, respectively. We take $c=0.1$, $s=1$, $\alpha=0.2$, $t^{s}_{1}=0$, $t^{f}_{1}=t^{s}_{2}=20$, $t^{f}_{2}=40$, $\epsilon=10^{-8}\omega_{c}$, and (a) $r=0$, (b) $r=0.5$, and (c) $r=1$.}
\label{fig5}
\end{figure}
In Fig. \ref{fig5}, we display the time evolution of the classical and quantum correlations as a function of the scaled time $\omega_{c}t$. Ohmicity parameter is $s=1$ for all three plots and the second interaction is turned on at $t=20$, at which point we turn off the first interaction. Fig. \ref{fig5}(a) demonstrates that, for the initial state with $c=0.1$, a sudden transition occurs before the second interaction is turned on, when there exists no entanglement between the reservoirs, corresponding to $r=0$. Before this point, discord is completely unaffected by the dephasing noise. We can also observe that the decay of discord hastens when we turn on the second interaction at $t=20$. In Figs. \ref{fig5}(b) and \ref{fig5}(c), we consider the effects of environmental correlations, i.e., non-zero entanglement between the environments, as we take the squeezing parameters as $r=0.5$ and $r=1$, respectively. Our results indicate that as the initial amount of the entanglement between the two reservoirs increase, sudden transition occurs earlier. Thus, we conclude that, in case the ohmicity parameter is taken as $s=1$, having initial environmental correlations are disadvantageous for preserving the quantum discord in the open quantum system.
\begin{figure}[b]
\includegraphics[scale=0.45]{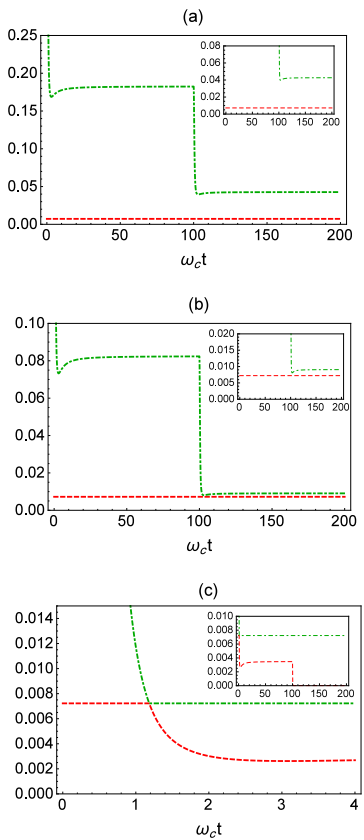}
\caption{(Color online) Dynamics of classical and quantum correlations as a function of the scaled time $\omega_{c}t$. The  The dotted-dashed green and dashed red lines represent the classical correlations and quantum discord, respectively. We take $c=0.1$, $s=2.5$, $\alpha=0.2$, $t^{s}_{1}=0$, $t^{f}_{1}=t^{s}_{2}=100$, $t^{f}_{2}=200$, and $\epsilon=10^{-8}\omega_{c}$. The squeezing parameter is taken as (a) $r=0$, (b) $r=0.5$, and (c) $r=1$.}
\label{fig6}
\end{figure}

We continue our study considering the case where each of the local interactions gives rise to non-Markovian dynamics. We assume that the ohmicity parameter is taken as $s=2.5$. In Fig. \ref{fig6}, we show the outcomes of our study for the same initial state with $c=0.1$ as in the previous case. Here, the first local interaction is switched off and the second one is switched on at the time $t^{s}_{2}=100$. Looking at Fig. \ref{fig6}(a), we confirm the existence of time-invariant quantum discord in the absence of initial entanglement between the two environments as $r=0$ in this plot. We also note that the classical correlations experience a sudden transition and suffer an abrupt decay. However, as we start to increase the amount of initial entanglement between the reservoirs in Fig. \ref{fig6}(b), we can observe that, even though time-invariant discord still persists, the sudden decay of classical correlations bring their value closer to the value of quantum discord. Finally, in Fig. \ref{fig6}(c), we see for strongly correlated environments ($r=1$) that classical correlations intersect with quantum discord in a very short time interval and time-invariant behaviour of quantum discord is lost. In fact, in this case, switching on the second interaction at $t^{s}_{2}=100$ completely destroys the remaining discord, while it does not affect the classical correlations, as shown in the inset.
\begin{figure}[t]
\includegraphics[scale=0.48]{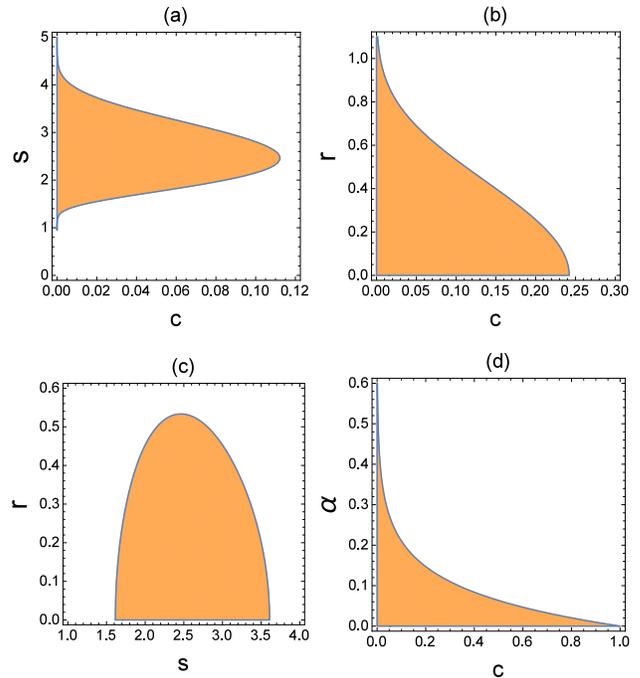}
\caption{(Color online) The orange shaded (dark-gray) regions show the range of various different parameters for which the time-invariant quantum discord exists. (a) $\alpha=0.2$ and $r=0.5$, (b) $\alpha=0.2$ and $s=2.5$, (c) $\alpha=0.2$ and $c=0.1$, and (d) $s=2.5$ and $r=0.5$.}
\label{fig7}
\end{figure}

Carefully analysing the Figs. \ref{fig5} and \ref{fig6}, we understand that the existence of initial  environmental correlations are generally detrimental for protecting quantum discord in open quantum systems. Specifically, as the amount of initial entanglement between the two reservoirs increase, we observe that in case we have frozen discord we can keep discord constant for shorter time intervals, and in case we have time-invariant discord we loose time-invariance since a sudden transition takes place. In order to gain a better understanding of how relevant system and model parameters affect the time-invariant nature of quantum discord, we present a detailed summary of our investigation in Fig. \ref{fig7}. Let us first assume that the coupling constants and squeezing parameter are fixed as $\alpha=0.2$, $r=0.5$. In Fig. \ref{fig7}(a), we plot the region in terms $s$ and $c$, where time-invariant discord can be observed. It is clear that the non-Markovian memory effects helps maintaining the quantum discord invariant throughout the dynamics since time-invariant discord exists when $1.5\lesssim s\lesssim 4$ for the majority of considered initial states. Now, we set $s=2.5$, $\alpha=0.2$ and display a plot for the region of time-invariant discord in terms of the strength of the initial entanglement between the environments $r$ and the initial state parameter $c$ in Fig. \ref{fig7}(b). As can be seen, time-invariant discord tend to disappear for the majority of the initial states as the initial correlations between the reservoirs increase. In fact, there exists no time-invariant entanglement once $r>1$. Next, we fix $c=0.1$, $\alpha=0.2$ in Fig. \ref{fig7}(c) and explore the region of time-invariant discord in terms of $r$ and $s$. It is straightforward to see that provided $r>0.5$, time-invariant discord can never be observed for any value of the ohmicitiy parameter $s$. Finally, we stress that the the shaded regions in Figs. \ref{fig7}(a), \ref{fig7}(b), and \ref{fig7}(c) naturally depend on the coupling constant $\alpha$, and in fact these regions will be widened for smaller values of $\alpha$. Fig. \ref{fig7}(d) shows the region of time-invariant discord in terms of the coupling constant $\alpha$ and the initial state parameter $c$ for $s=2.5$ and $r=0.5$. From this plot, we clearly see that the time-invariant discord favours the weak coupling regime. We emphasize that, outside the shaded regions in all plots, one always observes a sudden transition from classical to quantum decoherence and hence an eventual decay quantum discord.

\section{Conclusion}

In summary, we have performed a quite detailed analysis of the effects of some realistic features typically present in open quantum system models on the remarkable phenomena of frozen and time-invariant quantum discord. The motivation in studying such a problem comes from the critical significance of preserving the genuine quantum correlations in open quantum systems as they serve as a resource for various tasks in quantum information science. In particular, we have examined the consequences of dephasing, dissipation and heating reservoirs both individually and collectively for the occurrence of frozen and time-invariant discord. For the same purpose, we have also explored the influence of initial correlations between the two environments.

For instance, we have shown the existence of time-invariant discord for pure dephasing reservoirs at the high temperature limit provided we work in the weak coupling regime. Moreover, for dissipation/heating reservoirs, we have demonstrated that no sudden transition and thus frozen discord exists, but almost time-invariant discord can be observed with the proper choice of the detuning parameter that gives rise to memory effects. Also, considering a general model including the effects of both dephasing and dissipation/heating reservoirs, we have found out that the occurrence of time-invariant discord in this case is critically related with the existence of memory effects in the dephasing dynamics. Lastly, taking into account the possible initial correlations between two purely dephasing reservoirs, we have proved with our extensive analysis that the existence of such initial environmental correlations is in general quite detrimental for protecting and maintaining the genuine quantum correlations in open quantum systems.

\section*{Acknowledgements}
This work has been supported by the University of Kurdistan. F. T. T. thanks R. Yousefjani for the useful discussions. G. K. is grateful to Sao Paulo Research Foundation (FAPESP) for the support given under the grant number 2012/18558-5. S. M. also acknowledges financial support from the Academy of Finland (Project No. 287750).


\begin{thebibliography}{}
\bibitem{nielsenbook} M. A. Nielsen and I. L. Chuang, \textit{Quantum Computation and Quantum Information} (Cambridge University Press, Cambridge, 2000).
\bibitem{entreview} R. Horodecki, P. Horodecki, M. Horodecki, and K. Horodecki, Rev. Mod. Phys. \textbf{81}, 865 (2009).
\bibitem{discordreview} K. Modi, A. Brodutch, H. Cable, T. Paterek, and V. Vedral, Rev. Mod. Phys. \textbf{84}, 1655 (2012).
\bibitem{openbook} H. -P. Breuer and F. Petruccione, \textit{The Theory of Open Quantum Systems} (Oxford University Press, Oxford, 2002).
\bibitem{dd1} L. Viola and S. Lloyd, Phys. Rev. A \textbf{58}, 2733 (1998).
\bibitem{dd2} L. Viola, E. Knill and S. Lloyd, Phys. Rev. Lett. \textbf{82}, 2417 (1998).
\bibitem{dd3} C. Addis, G. Karpat, and S. Maniscalco, Phys. Rev. A \textbf{92}, 062109 (2015)
\bibitem{frozthry1} L. Mazzola, J. Piilo, and S. Maniscalco, Phys. Rev. Lett. \textbf{104}, 200401 (2010).
\bibitem{frozexp1} J. -S. Xu, X. -Y. Xu, C. -F. Li, C. -J. Zhang, X. -B. Zou, and G. -C. Guo, Nature Commun. \textbf{1}, 7 (2010).
\bibitem{frozexp2} R. Auccaise, L. C. Celeri, D. O. Soares-Pinto, E. R. deAzevedo, J. Maziero, A. M. Souza, T. J. Bonagamba, R. S. Sarthour, I. S. Oliveira, and R. M. Serra, Phys. Rev. Lett. \textbf{107}, 140403 (2011).
\bibitem{frozthry2} L. Mazzola, J. Piilo, and S. Maniscalco, Int. J. Quantum Inf. \textbf{9}, 981 (2011).
\bibitem{invdisc} P. Haikka, T. H. Johnson, and S. Maniscalco, Phys. Rev. A. \textbf{87}, 010103(R)(2013).
\bibitem{disc1} L. Henderson and V. Vedral, J. Phys. A \textbf{34}, 6899 (2001).
\bibitem{disc2} H. Ollivier and W. H. Zurek, Phys. Rev. Lett. \textbf{88}, 017901(2001).
\bibitem{discformula} F. F. Fanchini, T. Werlang, C. A. Brasil, L. G. E. Arruda, and A. O. Caldeira, Phys. Rev. A. \textbf{81}, 052107 (2010).
\bibitem{model1} A. Smirne, J. Kolodynski, S. F. Huelga, and R. Demkowicz-Dobrzanski, Phys. Rev. Lett. \textbf{116}, 120801 (2016).
\bibitem{model11} J. Lankinen, H. Lyyra, B. Sokolov, J. Teittinen, B. Ziaei, and S. Maniscalco, Phys. Rev. A.
\textbf{93}, 052103 (2016).
\bibitem{coupdetun} J. -G. Li, J. Zou, and B. Shao, Phys. Rev. A \textbf{81}, 062124 (2010).
\bibitem{nonmarkov} H.-P. Breuer, E. M. Laine, and J. Piilo, Phys. Rev. Lett. \textbf{103}, 210401 (2009).
\bibitem{model2}S. Wißmann, and H.-P. Breuer, Phys. Rev. A \textbf{90}, 032117 (2014).
\bibitem{fock} G. Giedke, M. M. Wolf, O. Kr\"{u}ger, R. F.Werner, and J. I. Cirac, Phys. Rev. Lett. \textbf{91}, 107901 (2003).
\bibitem{nonzdisc1}P. Giorda and M. G. A. Paris, Phys. Rev. Lett. \textbf{105}, 020503 (2010).
\bibitem{nonzdisc2}G. Adesso and A. Datta, Phys. Rev. Lett. \textbf{105}, 030501 (2010).
\end{thebibliography}
\end{document}